\documentclass[aps,prl,twocolumn,superscriptaddress,showpacs,preprintnumbers,amsmath,amssymb]{revtex4}
\usepackage{lipsum}
\usepackage{amsmath}
\usepackage{graphicx} 
\usepackage{dcolumn}  
\usepackage{color}
\RequirePackage{xspace}
\usepackage{relsize}
\usepackage{multirow}
\usepackage{subfigure}
\usepackage{threeparttable}
\usepackage{mathtools}
\usepackage{verbatim}
\usepackage[columnwise]{lineno}

\begin{document}


\preprint{\vbox{ \hbox{   }
    \hbox{BELLE-CONF-1607}
                        \hbox{Author: I.S.Lee}
                        \hbox{Committee: B. Pal (chair),}
                        \hbox{Y. Guan, J. Yamaoka }
}}

\title{ \quad\\[2.0cm] Evidence for the decay $B^{0}\to \eta \eta$ }

\noaffiliation
\affiliation{Aligarh Muslim University, Aligarh 202002}
\affiliation{University of the Basque Country UPV/EHU, 48080 Bilbao}
\affiliation{Beihang University, Beijing 100191}
\affiliation{University of Bonn, 53115 Bonn}
\affiliation{Budker Institute of Nuclear Physics SB RAS, Novosibirsk 630090}
\affiliation{Faculty of Mathematics and Physics, Charles University, 121 16 Prague}
\affiliation{Chiba University, Chiba 263-8522}
\affiliation{Chonnam National University, Kwangju 660-701}
\affiliation{University of Cincinnati, Cincinnati, Ohio 45221}
\affiliation{Deutsches Elektronen--Synchrotron, 22607 Hamburg}
\affiliation{University of Florida, Gainesville, Florida 32611}
\affiliation{Department of Physics, Fu Jen Catholic University, Taipei 24205}
\affiliation{Justus-Liebig-Universit\"at Gie\ss{}en, 35392 Gie\ss{}en}
\affiliation{Gifu University, Gifu 501-1193}
\affiliation{II. Physikalisches Institut, Georg-August-Universit\"at G\"ottingen, 37073 G\"ottingen}
\affiliation{SOKENDAI (The Graduate University for Advanced Studies), Hayama 240-0193}
\affiliation{Gyeongsang National University, Chinju 660-701}
\affiliation{Hanyang University, Seoul 133-791}
\affiliation{University of Hawaii, Honolulu, Hawaii 96822}
\affiliation{High Energy Accelerator Research Organization (KEK), Tsukuba 305-0801}
\affiliation{J-PARC Branch, KEK Theory Center, High Energy Accelerator Research Organization (KEK), Tsukuba 305-0801}
\affiliation{Hiroshima Institute of Technology, Hiroshima 731-5193}
\affiliation{IKERBASQUE, Basque Foundation for Science, 48013 Bilbao}
\affiliation{University of Illinois at Urbana-Champaign, Urbana, Illinois 61801}
\affiliation{Indian Institute of Science Education and Research Mohali, SAS Nagar, 140306}
\affiliation{Indian Institute of Technology Bhubaneswar, Satya Nagar 751007}
\affiliation{Indian Institute of Technology Guwahati, Assam 781039}
\affiliation{Indian Institute of Technology Madras, Chennai 600036}
\affiliation{Indiana University, Bloomington, Indiana 47408}
\affiliation{Institute of High Energy Physics, Chinese Academy of Sciences, Beijing 100049}
\affiliation{Institute of High Energy Physics, Vienna 1050}
\affiliation{Institute for High Energy Physics, Protvino 142281}
\affiliation{Institute of Mathematical Sciences, Chennai 600113}
\affiliation{INFN - Sezione di Torino, 10125 Torino}
\affiliation{Advanced Science Research Center, Japan Atomic Energy Agency, Naka 319-1195}
\affiliation{J. Stefan Institute, 1000 Ljubljana}
\affiliation{Kanagawa University, Yokohama 221-8686}
\affiliation{Institut f\"ur Experimentelle Kernphysik, Karlsruher Institut f\"ur Technologie, 76131 Karlsruhe}
\affiliation{Kavli Institute for the Physics and Mathematics of the Universe (WPI), University of Tokyo, Kashiwa 277-8583}
\affiliation{Kennesaw State University, Kennesaw, Georgia 30144}
\affiliation{King Abdulaziz City for Science and Technology, Riyadh 11442}
\affiliation{Department of Physics, Faculty of Science, King Abdulaziz University, Jeddah 21589}
\affiliation{Korea Institute of Science and Technology Information, Daejeon 305-806}
\affiliation{Korea University, Seoul 136-713}
\affiliation{Kyoto University, Kyoto 606-8502}
\affiliation{Kyungpook National University, Daegu 702-701}
\affiliation{\'Ecole Polytechnique F\'ed\'erale de Lausanne (EPFL), Lausanne 1015}
\affiliation{P.N. Lebedev Physical Institute of the Russian Academy of Sciences, Moscow 119991}
\affiliation{Faculty of Mathematics and Physics, University of Ljubljana, 1000 Ljubljana}
\affiliation{Ludwig Maximilians University, 80539 Munich}
\affiliation{Luther College, Decorah, Iowa 52101}
\affiliation{University of Maribor, 2000 Maribor}
\affiliation{Max-Planck-Institut f\"ur Physik, 80805 M\"unchen}
\affiliation{School of Physics, University of Melbourne, Victoria 3010}
\affiliation{Middle East Technical University, 06531 Ankara}
\affiliation{University of Miyazaki, Miyazaki 889-2192}
\affiliation{Moscow Physical Engineering Institute, Moscow 115409}
\affiliation{Moscow Institute of Physics and Technology, Moscow Region 141700}
\affiliation{Graduate School of Science, Nagoya University, Nagoya 464-8602}
\affiliation{Kobayashi-Maskawa Institute, Nagoya University, Nagoya 464-8602}
\affiliation{Nara University of Education, Nara 630-8528}
\affiliation{Nara Women's University, Nara 630-8506}
\affiliation{National Central University, Chung-li 32054}
\affiliation{National United University, Miao Li 36003}
\affiliation{Department of Physics, National Taiwan University, Taipei 10617}
\affiliation{H. Niewodniczanski Institute of Nuclear Physics, Krakow 31-342}
\affiliation{Nippon Dental University, Niigata 951-8580}
\affiliation{Niigata University, Niigata 950-2181}
\affiliation{University of Nova Gorica, 5000 Nova Gorica}
\affiliation{Novosibirsk State University, Novosibirsk 630090}
\affiliation{Osaka City University, Osaka 558-8585}
\affiliation{Osaka University, Osaka 565-0871}
\affiliation{Pacific Northwest National Laboratory, Richland, Washington 99352}
\affiliation{Panjab University, Chandigarh 160014}
\affiliation{Peking University, Beijing 100871}
\affiliation{University of Pittsburgh, Pittsburgh, Pennsylvania 15260}
\affiliation{Punjab Agricultural University, Ludhiana 141004}
\affiliation{Research Center for Electron Photon Science, Tohoku University, Sendai 980-8578}
\affiliation{Research Center for Nuclear Physics, Osaka University, Osaka 567-0047}
\affiliation{Theoretical Research Division, Nishina Center, RIKEN, Saitama 351-0198}
\affiliation{RIKEN BNL Research Center, Upton, New York 11973}
\affiliation{Saga University, Saga 840-8502}
\affiliation{University of Science and Technology of China, Hefei 230026}
\affiliation{Seoul National University, Seoul 151-742}
\affiliation{Shinshu University, Nagano 390-8621}
\affiliation{Showa Pharmaceutical University, Tokyo 194-8543}
\affiliation{Soongsil University, Seoul 156-743}
\affiliation{University of South Carolina, Columbia, South Carolina 29208}
\affiliation{Stefan Meyer Institute for Subatomic Physics, Vienna 1090}
\affiliation{Sungkyunkwan University, Suwon 440-746}
\affiliation{School of Physics, University of Sydney, New South Wales 2006}
\affiliation{Department of Physics, Faculty of Science, University of Tabuk, Tabuk 71451}
\affiliation{Tata Institute of Fundamental Research, Mumbai 400005}
\affiliation{Excellence Cluster Universe, Technische Universit\"at M\"unchen, 85748 Garching}
\affiliation{Department of Physics, Technische Universit\"at M\"unchen, 85748 Garching}
\affiliation{Toho University, Funabashi 274-8510}
\affiliation{Tohoku Gakuin University, Tagajo 985-8537}
\affiliation{Department of Physics, Tohoku University, Sendai 980-8578}
\affiliation{Earthquake Research Institute, University of Tokyo, Tokyo 113-0032}
\affiliation{Department of Physics, University of Tokyo, Tokyo 113-0033}
\affiliation{Tokyo Institute of Technology, Tokyo 152-8550}
\affiliation{Tokyo Metropolitan University, Tokyo 192-0397}
\affiliation{Tokyo University of Agriculture and Technology, Tokyo 184-8588}
\affiliation{University of Torino, 10124 Torino}
\affiliation{Toyama National College of Maritime Technology, Toyama 933-0293}
\affiliation{Utkal University, Bhubaneswar 751004}
\affiliation{Virginia Polytechnic Institute and State University, Blacksburg, Virginia 24061}
\affiliation{Wayne State University, Detroit, Michigan 48202}
\affiliation{Yamagata University, Yamagata 990-8560}
\affiliation{Yonsei University, Seoul 120-749}
  \author{A.~Abdesselam}\affiliation{Department of Physics, Faculty of Science, University of Tabuk, Tabuk 71451} 
  \author{I.~Adachi}\affiliation{High Energy Accelerator Research Organization (KEK), Tsukuba 305-0801}\affiliation{SOKENDAI (The Graduate University for Advanced Studies), Hayama 240-0193} 
  \author{K.~Adamczyk}\affiliation{H. Niewodniczanski Institute of Nuclear Physics, Krakow 31-342} 
  \author{H.~Aihara}\affiliation{Department of Physics, University of Tokyo, Tokyo 113-0033} 
  \author{S.~Al~Said}\affiliation{Department of Physics, Faculty of Science, University of Tabuk, Tabuk 71451}\affiliation{Department of Physics, Faculty of Science, King Abdulaziz University, Jeddah 21589} 
  \author{K.~Arinstein}\affiliation{Budker Institute of Nuclear Physics SB RAS, Novosibirsk 630090}\affiliation{Novosibirsk State University, Novosibirsk 630090} 
  \author{Y.~Arita}\affiliation{Graduate School of Science, Nagoya University, Nagoya 464-8602} 
  \author{D.~M.~Asner}\affiliation{Pacific Northwest National Laboratory, Richland, Washington 99352} 
  \author{T.~Aso}\affiliation{Toyama National College of Maritime Technology, Toyama 933-0293} 
  \author{H.~Atmacan}\affiliation{Middle East Technical University, 06531 Ankara} 
% ...
\collaboration{The Belle Collaboration}

\clearpage

\begin{abstract}
We report a search for  $B^{0}\to \eta \eta $ with a data sample corresponding to an integrated luminosity of $698 \,{\rm fb}^{-1}$ containing $753 \times 10^{6}$ $B\bar{B}$ pairs collected at the $\Upsilon(4S)$ resonance with the Belle detector at the KEKB asymmetric-energy $e^{+}e^{-}$ collider. The branching fraction is measured to be $\mathcal{B}(B^{0} \to \eta \eta ) = (7.6^{+2.7 +1.4}_{-2.3 -1.6}) \times 10^{-7}$ at the level of 3.3 standard deviations above zero, which provides the first evidence for the decay $B^{0} \to \eta \eta$. 
\end{abstract}


\maketitle

\tighten

{\renewcommand{\thefootnote}{\fnsymbol{footnote}}}
\setcounter{footnote}{0}

\section{INTRODUCTION}

The $CP$ violation measurements using charmless hadronic decays of $B$ mesons are primarily important for testing the Standard Model (SM) and searching for physics beyond the SM.

\begin{figure}[h!]
 \centering

\subfigure[Tree diagram \label{fig:Feynman:a}]{\includegraphics[width = 4.0cm]{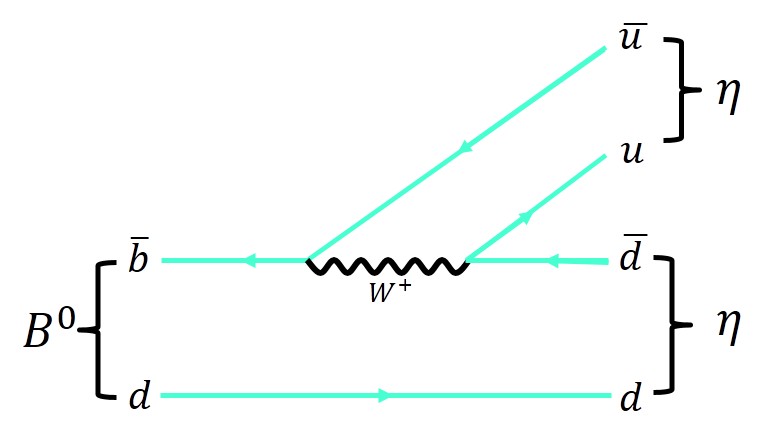}}  
\subfigure[Penguin diagram\label{fig:Feynman:b}]{\includegraphics[width = 4.0cm]{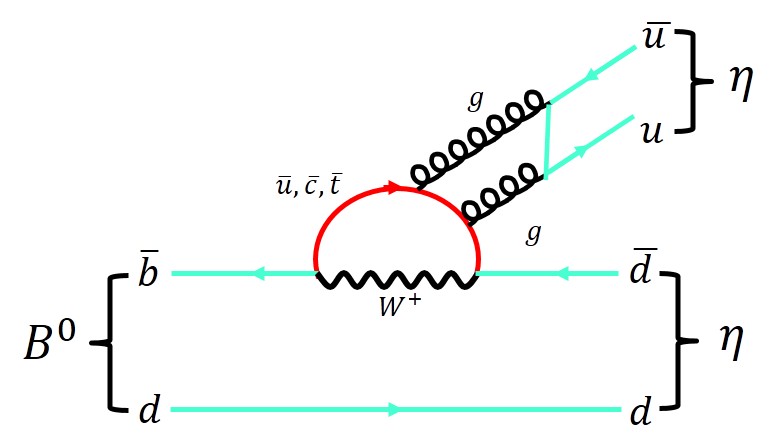}}
\label{fig:Feynman}
\caption{Diagrams for  $B^0 \to \eta \eta$ decays }
\end{figure}

 The $B^0 \to \eta \eta$ decay mode mainly proceeds via the $b\to u$ Cabibbo- and color-suppressed tree diagram and the $b\to d$ penguin diagram shown in Fig.~\ref{fig:Feynman:a} and Fig.~\ref{fig:Feynman:b}, respectively. 
The expected branching fraction of this decay mode is $(0.3 - 3.1)\times 10^{-6}$, estimated from the calculations based on QCD factorization~\cite{int12}, soft collinear effective theory~\cite{int13}, SU(3) flavor symmetry~\cite{int14} and flavor U(3) symmetry~\cite{int15}.

This mode plays an important role in improving the flavor SU(3) calculations of $|S_{ccs}-S_{f}|$, where the final state $f$ is $\eta^{'}K$ or $\phi K$, the $CP$-violating parameter $S_{f} \sim \sin{2\phi_{1}}$ is measured in the time-dependent analysis~\cite{int16}, and the $CP$-violating parameter $S_{c\bar{c}s}$ is measured in the Cabibbo-Kobayashi-Maskawa(CKM)-favored $b\to c\bar{c}s$ decays. 
The bound on this difference can be improved by more precise measurements of the branching fraction of $B^0 \to \eta \eta$~\cite{int10, int11}.

This mode has been studied by Belle and BABAR~\cite{int22, int23}. 
The best previous upper limit on this branching fraction is $\mathcal{B}( B^0 \to \eta \eta ) <1.0 \times 10^{-6}$ at $90 \%$ confidence level (CL)~\cite{int23}. 

We update our previous result~\cite{int22} using the full data set of the Belle experiment running on the $\Upsilon(4S)$ resonance at the KEKB asymmetric-energy $e^{+}e^{-}$ collider~\cite{int24}.
This data set corresponds to $753 \times 10^{6}$ $B\bar{B}$ pairs, which is a factor of five larger than in the previous Belle study. 
\\

\section{DETECTOR AND DATASET}
The Belle detector is a large-solid-angle magnetic
spectrometer that consists of a silicon vertex detector (SVD),
a 50-layer central drift chamber (CDC), an array of
aerogel threshold Cherenkov counters (ACC),  
a barrel-like arrangement of time-of-flight
scintillation counters (TOF), and an electromagnetic calorimeter
comprised of CsI(Tl) crystals (ECL) located inside 
a superconducting solenoid coil that provides a 1.5~T
magnetic field.  An iron flux-return located outside of
the coil is instrumented to detect $K_L^0$ mesons and to identify
muons (KLM).  The detector
is described in detail elsewhere~~\cite{int24}.
Two inner detector configurations were used. A 2.0 cm beampipe
and a 3-layer silicon vertex detector was used for the first sample
of $133 \times 10^6 B\bar{B}$ pairs, while a 1.5 cm beampipe, a 4-layer
silicon detector and a small-cell inner drift chamber were used to record the remaining $620 \times 10^6 B\bar{B}$ pairs~\cite{svd2}.  

\section{RECONSTRUCTION}
The $B^{0}\to \eta \eta$ candidate is reconstructed from the sub-decay channels of $\eta \to \gamma \gamma$$\,$($\eta_{\gamma\gamma}$) and $\eta \to \pi^{+}\pi^{-}\pi^{0}$$\,$($\eta_{3\pi}$).  

Photons used for $\eta\to\gamma\gamma$ and $\pi^{0} \to \gamma\gamma$ are required to have an energy greater than $50\,(100)\textrm{\,MeV}$ in the barrel (end-cap) region of the ECL~\cite{int24}. 
The ECL timing information, which measures the time of energy deposit relative to the beam-collision time, is used to reject out-of-time photons. 

 Charged tracks are required to have a transverse momentum greater than $0.1\,\textrm{GeV}/c$ and an impact parameter with respect to the interaction point of less than $0.3\textrm{\,cm}$ in the $r-\phi$ plane and $3.0\textrm{\,cm}$ along the $z$ axis, which is opposite the $e^{+}$ beam.
Charged pions are identified using information obtained from the CDC$\,$($dE/dx$), the TOF, and the ACC$\,$(number of photoelectrons). This information is combined to form a likelihood $\mathcal{L}$ for hadron identification$\,$(PID). 
We require that charged pions satisfy $\mathcal{L}_{K}/(\mathcal{L}_{\pi}+\mathcal{L}_{K}) < 0.4$, where $\mathcal{L}_{K}\,$($\mathcal{L}_{\pi}$) denotes the likelihood for a track with the kaon (pion) hypothesis.
 
A $\pi^{0}$ candidate is required to have a $\gamma\gamma$ invariant mass between $117$ and $155 \textrm{\,MeV}/c^{2}$, which corresponds to $\pm 3\sigma$ around the nominal $\pi^{0}$ mass~\cite{PDG}. 
To improve the $\pi^{0}$ momentum resolution, we perform a mass-constrained fit and require that the resulting goodness-of-fit parameter ($\chi^{2}$) be less than 50. 

Photons that are not included in the final set of $\pi^0$ candidates are combined to form $\eta_{\gamma\gamma}$ candidates.
To reduce combinatorial background from low-energy photons, $\eta_{\gamma\gamma}$ candidates are required to have $\cos{\theta_{\eta}}<0.9$, where $\theta_{\eta}$ is the angle in the $\eta$ rest frame between the directions of the $B$ meson and the more energetic photon.
The $\eta_{3\pi}$ candidates are reconstructed by combining two oppositely charged pion candidates with a $\pi^{0}$ candidate.
We require the invariant mass of $\eta_{\gamma\gamma}$$\,$($\eta_{3\pi}$) candidates be in the range $476-579 \,$($527-568$)$\textrm{\,MeV}/c^{2}$, which corresponds to $\pm 2.5\sigma \,$($\pm 3.0 \sigma$) around the nominal $\eta$ mass~\cite{PDG}.
For each such $\eta$ candidate, a mass-constrained fit is performed to improve the momentum resolution; the resulting $\chi^{2}$ of the $\eta_{\gamma\gamma}\,$($\eta_{3\pi}$) candidates must be less than 50$\,$(200).

 For each $B^0 \to \eta \eta$ candidate, we define two kinematic variables to distinguish signal from background: these are the energy difference ($\small{\Delta E \equiv E_{B} -E_{\rm beam}}$) and the beam-constrained mass ($\tiny M_{\rm bc}\equiv \sqrt{E^{2}_{\rm beam}-\left|\vec{p}_{B}\right|^{2}c^{2}}/c^{2}$), where $E_{\rm beam}$ and $E_{B}\,$($\vec{p}_{B}$) are the beam energy and the energy$\,$(momentum) of the $B$ meson candidate, respectively, in the $e^{+}e^{-}$ center-of-mass$\,$(CM) frame. 
We retain $B$ candidates satisfying  $-0.3\, \textrm{GeV} < \Delta E< 0.2 \, \textrm{GeV}$ and $M_{\rm bc} > 5.25\textrm{\,GeV/}c^{2}$.

\section{BACKGROUND SUPPRESSION}

The large background arising from continuum $e^{+}e^{-}\to q\bar{q}$\,$(q=u, d, s, c)$ process is the dominant background here.  
To suppress this background, a neural network~\cite{int25} is employed by combining the following four quantities: the event-shape variables formed from 16 modified Fox-Wolfram moments~\cite{KSFW}, the cosine of the angle between the flight direction of the $B$ candidate and $z$ axis in the $e^{+}e^{-}$ CM frame, the tagging information of the flavor~\cite{TaggingNIM}, and the cosine of the angle between the thrust axes~\cite{thrust} in the $e^{+}e^{-}$ CM frame of the signal-$B$ and the other-$B$ candidates. 
The training and optimization of this neural network are accomplished with signal and continuum Monte Carlo (MC) simulated events. The signal MC sample is generated with EvtGen~\cite{evtgen}, taking the final-state radiation into account via PHOTOS~\cite{photos}.
After training, independent samples are used to test the neural network performance.  
The neural network output ($C_{\mathit{NB}}$) for an event ranges from $-1$ to $+1$; a value near $+1\,$($-1$) is more likely signal$\,$(continuum).

We require $C_{\mathit{NB}}>-0.8$ to suppress the continuum background. 
This requirement preserves approximately $97.7\%$, $97.6\%$ and $97.2\%$ of the signal while suppressing $68.3\%$, $64.5\%$, and $58.5\%$ of the continuum background in $\eta_{\gamma\gamma}\eta_{\gamma\gamma}$, $\eta_{\gamma\gamma}\eta_{3\pi}$ and $\eta_{3\pi}\eta_{3\pi}$, respectively.
As the remainder of the $C_{\mathit{NB}}$ distribution has a strong peak that falls off rapidly below 1 for signal, we use a transformed quantity to improve its modeling with an analytic shape:
\begin{eqnarray}
  C^{'}_{\mathit{NB}} = \log{\frac{C_{\mathit{NB}}-C^{\rm min}_{\mathit{NB}}}{C^{\rm max}_{\mathit{NB}}-C_{\mathit{NB}}}},
\end{eqnarray}
where $C^{\rm min}_{\mathit{NB}}= -0.8$ and $C^{\rm max}_{\mathit{NB}}$ is the maximum value of $C_{\mathit{NB}}$ obtained from a large sample of signal MC decays. 

After applying all selection criteria, the average number of signal candidates per event is 1.08, 1.10 and 1.13 for $\eta_{\gamma\gamma}\eta_{\gamma\gamma}$, $\eta_{\gamma\gamma}\eta_{3\pi}$ and $\eta_{3\pi}\eta_{3\pi}$, respectively, in the signal MC.  
We choose the candidate having the smallest value for the sum of the $\chi^{2}$ values from the two $\eta$ mass-constrained fits. 
We refer to the right-combination$\,$(RC) as the correctly-reconstructed $B$ meson decays and the self-crossfeed$\,$(SCF) as the misreconstructed signal components.
 MC simulation show that the SCF fraction is $6.8\%$, $9.3\%$ and $13.4\%$ in $\eta_{\gamma\gamma}\eta_{\gamma\gamma}$, $\eta_{\gamma\gamma}\eta_{3\pi}$ and $\eta_{3\pi}\eta_{3\pi}$, respectively.

\section{FIT MODEL}

 \begin{table}[t!]
   \caption{\small List of PDFs used to model $\Delta E$, $M_{\rm bc}$ and $C^{'}_{\mathit{NB}}$ distributions for various event categories. G, AG, CB, NV, i-th CC, ARG and HistPDF stand for Gaussian,asymmetric Gaussian,Crystal Ball~\cite{CB}, Novosibirsk~\cite{NV}, i-th Chebyshev polynomial, ARGUS function~\cite{ARG} and histogram, respectively. $N_{i}$ is the yields of signal or background components.}
\begin{center}
\begin{tabular}{c|c|c|c|c|c}
\hline \hline
\multicolumn{2}{c|}{Fit variables} & Yield & $\Delta E$ & $M_{bc}$ & $C'_{\mathit{NB}}$ \\ \hline
\multirow{2}{*}{Signal MC} & RC &\multirow{2}{*}{$N_{\rm sig}$} & CB + G &  CB & AG + AG \\ \cline{2-2} \cline{4-6}
                                            & SCF &                                       & AG + 1st CC & NV + ARG & AG + AG \\ \hline
\multicolumn{2}{c|}{Continuum $q\bar{q}$} & $N_{qq}$ & 1st CC  & ARG & NV+ G \\ \hline
\multicolumn{2}{c|}{$b \to c$} & $N_{bb}$ & HistPDF  & HistPDF & G \\ \hline
\multicolumn{2}{c|}{$b \to u, d, s$ } & $N_{rare}$ & 2nd CC  & NV & G \\ \hline
\end{tabular}
\end{center}

\label{table:pdf222333}
\end{table}

The branching fraction of $B^{0}\to \eta \eta $ is obtained using a simultaneous fit to the $\eta_{\gamma\gamma}\eta_{\gamma\gamma}$, $\eta_{\gamma\gamma}\eta_{3\pi}$ and $\eta_{3\pi}\eta_{3\pi}$ decay channels.
In this fit, the branching fraction is determined by an unbinned extended maximum likelihood (ML) fit to the distributions of $\Delta E$, $M_{\rm bc}$ and $C^{'}_{\mathit{NB}}$. 
The branching fraction of this mode is obtained using a simultaneous fit to the $\eta_{\gamma\gamma}\eta_{\gamma\gamma}$, $\eta_{\gamma\gamma}\eta_{3\pi}$ and $\eta_{3\pi}\eta_{3\pi}$ decay channels.
The extended ML function is defined as 
\begin{eqnarray}
  \mathcal{L}_{\textrm{fit}} = e^{-\Sigma_{j}N_{j}}\cdot \prod_{i}^{M}\left( \sum_{j}N_{j} \mathcal{P}_{j}\left(\Delta E^{i}, M^{i}_{\mathrm{bc}},C^{' i}_{\mathit{NB}} \right) \right) ,
\label{likelihood}
\end{eqnarray}
where $\mathcal{P}_{j}\left(\Delta E, M_{\rm bc},C^{'}_{\mathit NB} \right)$ is the probability density function$\,$(PDF) of the signal or background component (specified by index $j$), $N_{j}$ is the fractional yield of this component for event $i$, and $M$ is the total number of events in the sample.  
The background components include continuum events, the $b\to c$ process, charmless $b \to u, d, s$ processes other than $B^{0}\to \pi^{0} \eta$, and $B^{0} \to \pi^{0}\eta$ (treated separately).
Compared to continuum, the other background processes are small and are modeled using MC simulations. 
The expected yields of $b \to c$ processes are 6, 6 and 2 events in $\eta_{\gamma\gamma}\eta_{\gamma\gamma}$, $\eta_{\gamma\gamma}\eta_{3\pi}$ and $\eta_{3\pi}\eta_{3\pi}$, respectively, after passing all selection criteria based on MC simulations.  
The charmless $b \to u, d, s$ processes (excluding $B^{0}\to\pi^{0}\eta$), while having larger expected yields, exhibit no peaking structure in the $\Delta E$ and $M_{\rm bc}$. 

We find that the correlations among the fit variables are small enough to ignore.
Thus, the three-dimensional PDF, $\mathcal{P}_{j}$ is expressed as the product of one-dimensional PDFs as 
\begin{eqnarray}
\mathcal{P}_{j} = \mathcal{P}_{j}(\Delta E) \cdot \mathcal{P}_{j}(M_{\rm bc}) \cdot \mathcal{P}_{j}(C_{\mathit{NB}}^{'}). 
\label{PDF}
\end{eqnarray}
Table \ref{table:pdf222333} lists the PDF shapes used to model $\Delta E$, $M_{\rm bc}$ and $C^{'}_{\mathit{NB}}$ for all components in the fit. 

We fix the parameters of the RC-signal PDF shapes to the values obtained from the signal MC.
Here, the peak positions and resolutions are adjusted according to data-MC differences observed in a high-statistics control sample of $B^{0}\to \overline{D}{}^{0}(\to K^{+}\pi^{-}\pi^{0}) \eta$ decays. 

 The continuum $q\bar{q}$ background PDF parameters that are allowed to vary are the slope of $\Delta E$, the shape of $M_{\rm bc}$ and the mean and width of the $C^{'}_{\mathit {NB}}$ Gaussian function. 
All of the other background PDF parameters are fixed based on MC simulation. 

The yields of signal and continuum $q\bar{q}$ are allowed to vary in the fit. 
The yields of the other backgrounds and the relative amount of SCF to RC signal are fixed. 
To test the stability of the fitter, we perform the fit to ensembles of 1000 pseudoexperiments using the extracted fitted yields from data and random samples of events for each component chosen from the simulated MC samples. 
We observe a fit bias of $1.7\%$ for $\mathcal{B}$, which is corrected and assigned as a systematic uncertainty.

\begin{figure*}[t!]
 \centering
 \subfigure[$B^{0}\to\eta_{\gamma\gamma}\eta_{\gamma\gamma}$]{  \includegraphics[width = 5cm]{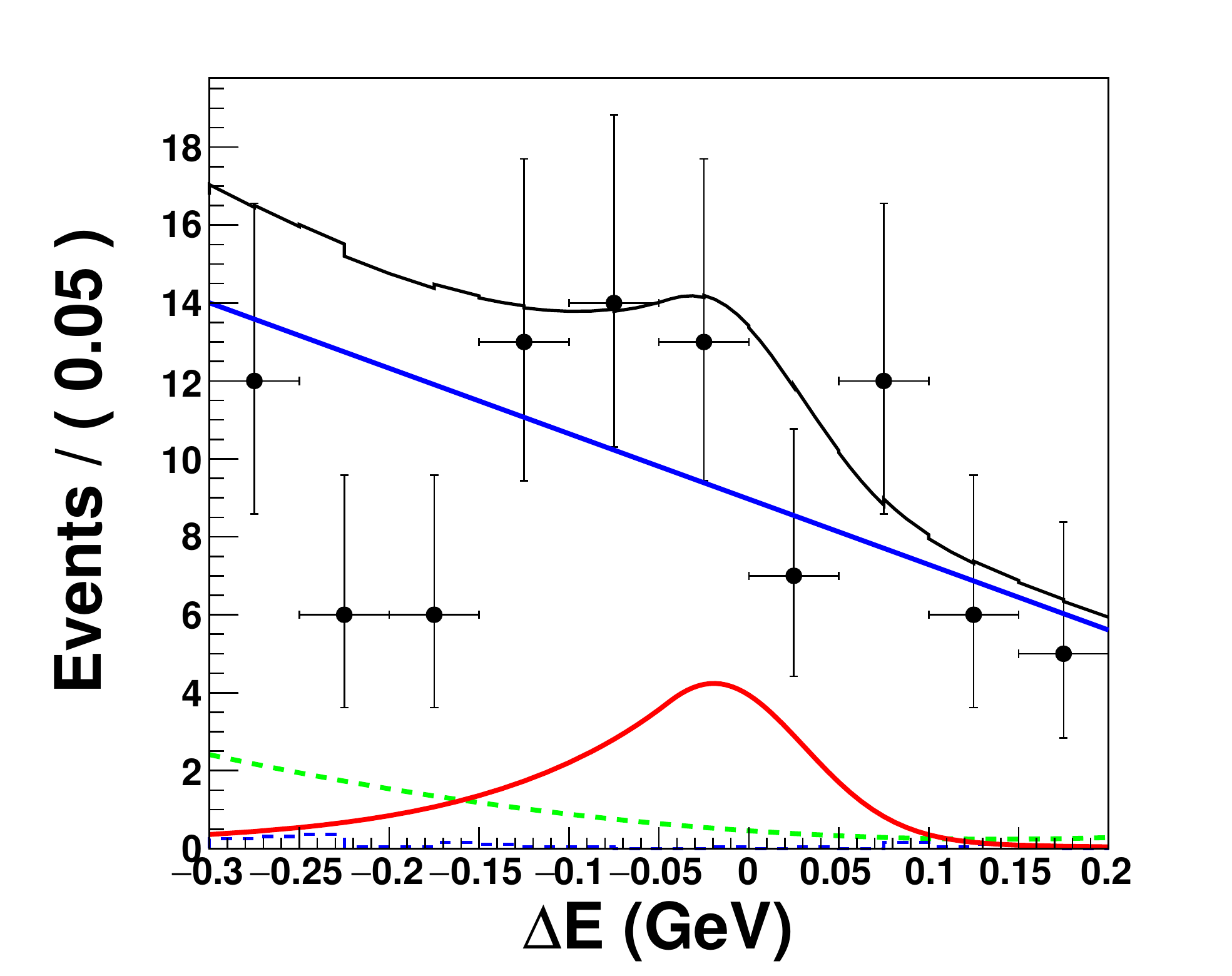} 
 \includegraphics[width = 5cm]{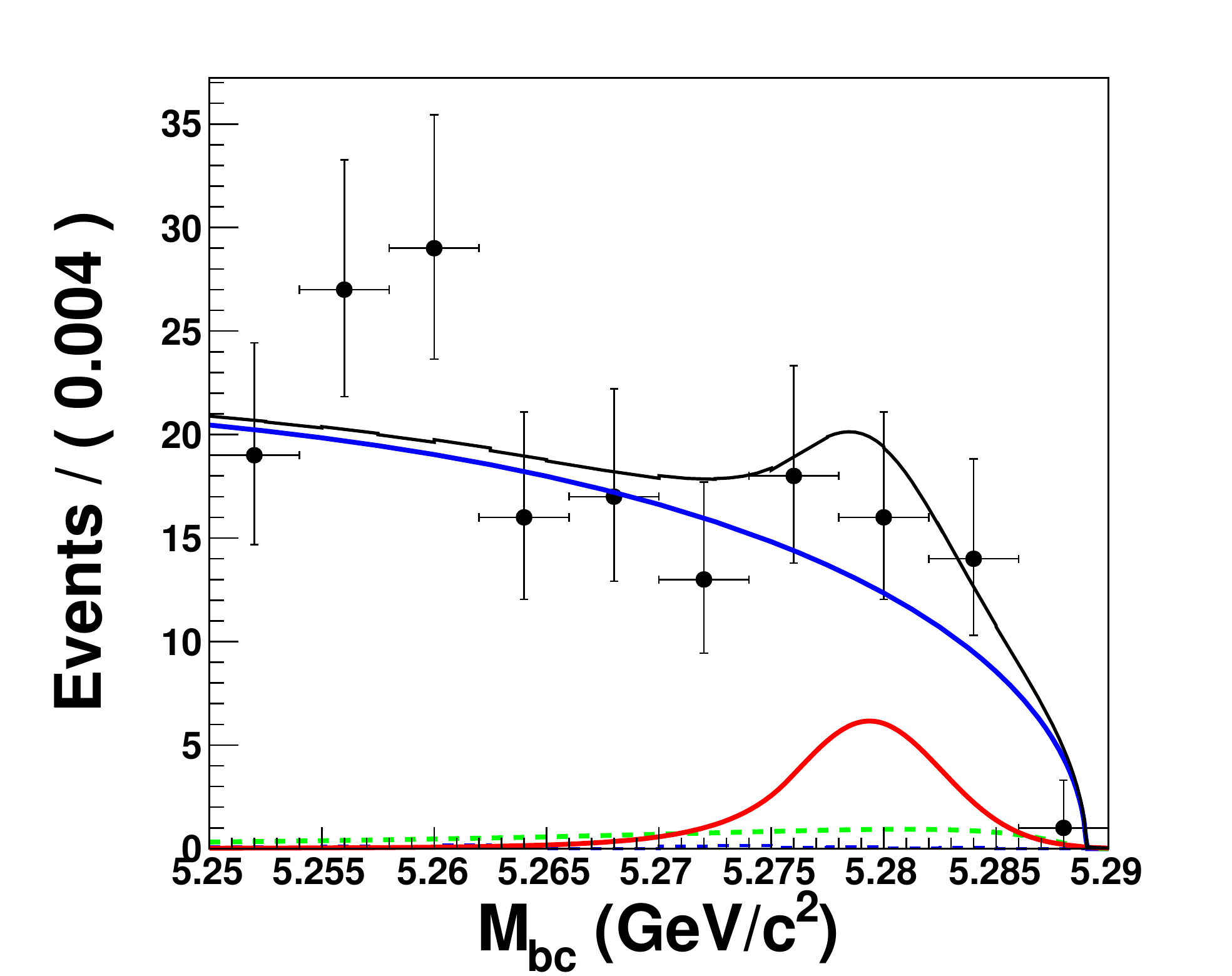}
 \includegraphics[width = 5cm]{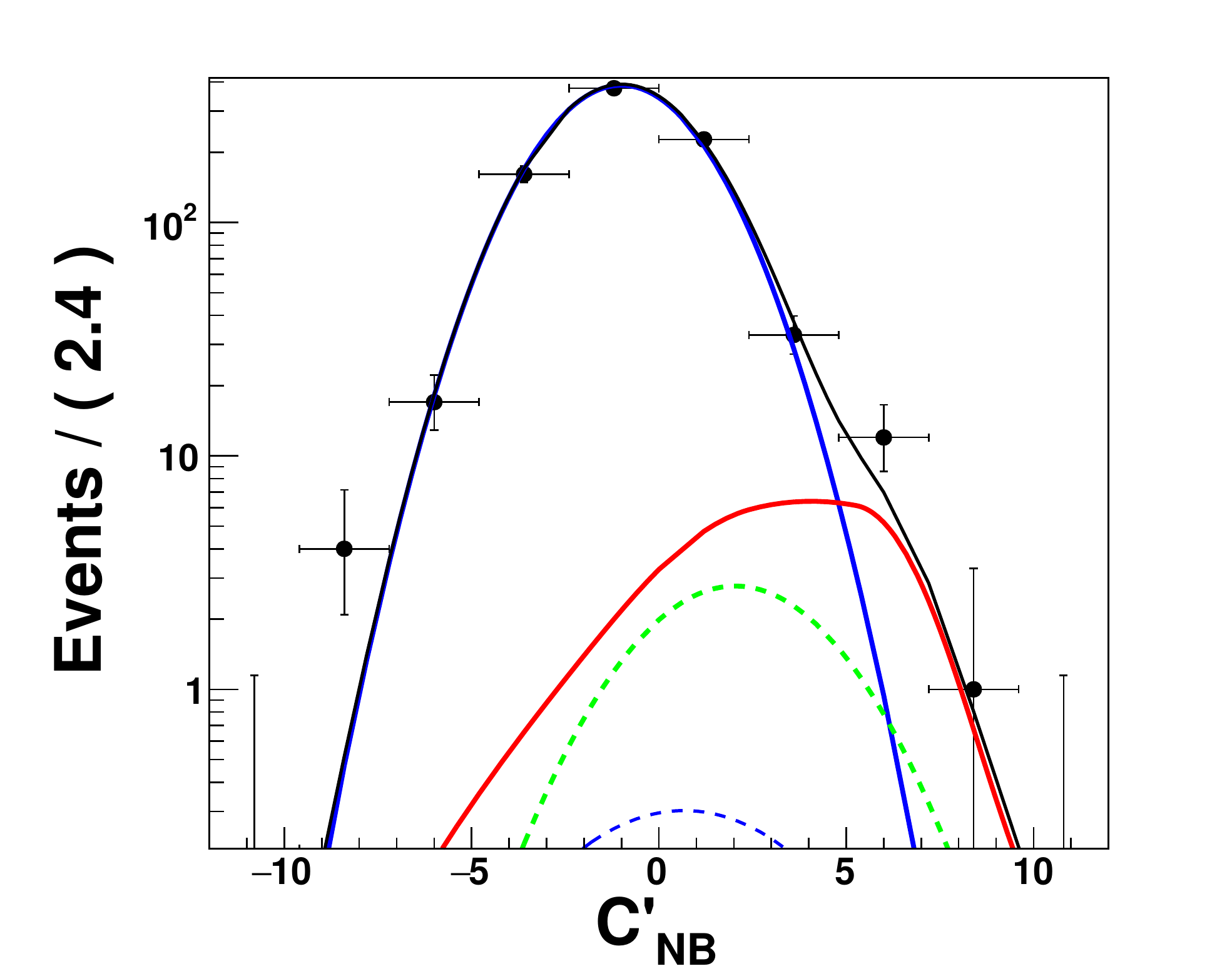}
}
 \subfigure[$B^{0}\to\eta_{\gamma\gamma}\eta_{3\pi}$]{  \includegraphics[width = 5cm]{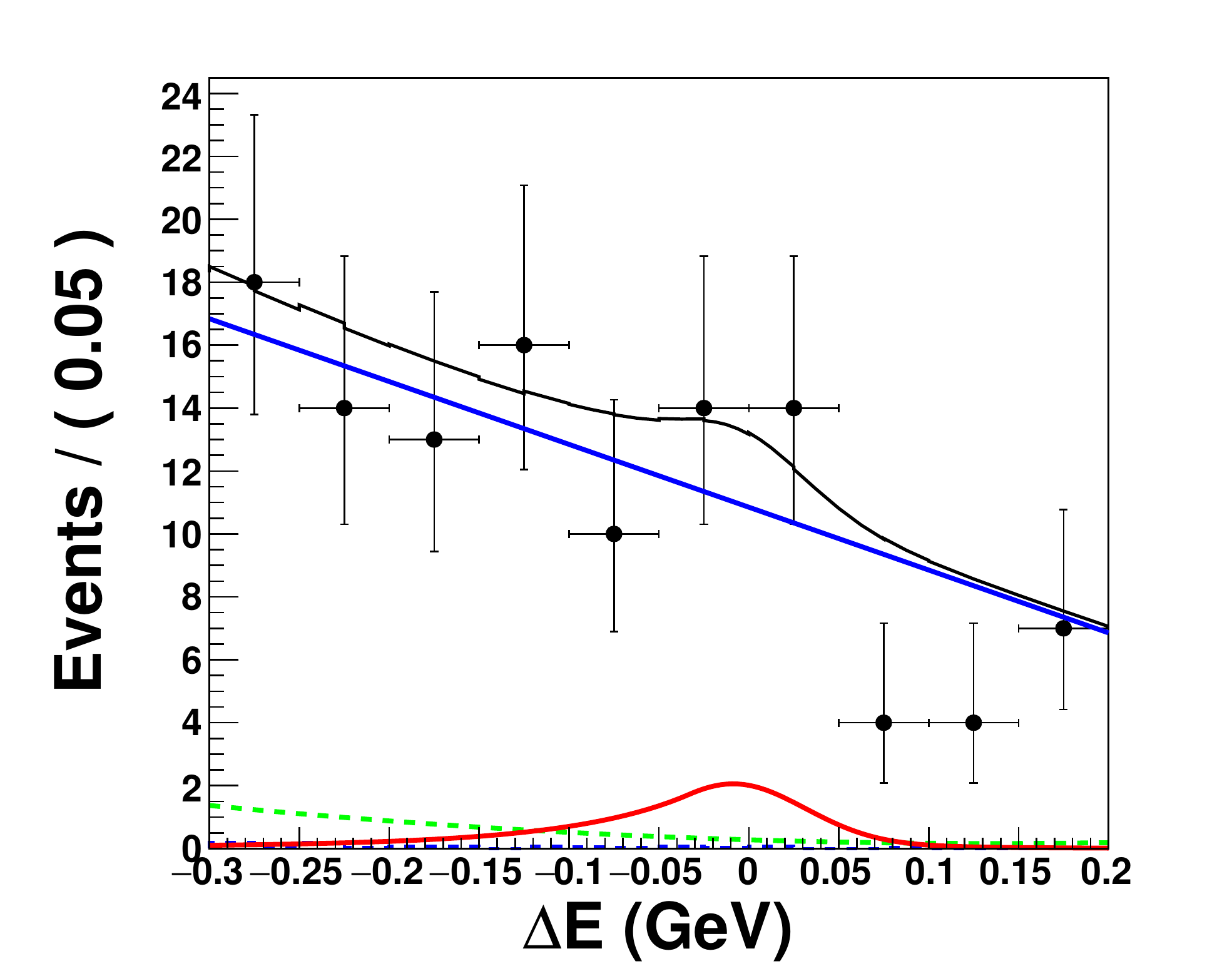} 
 \includegraphics[width = 5cm]{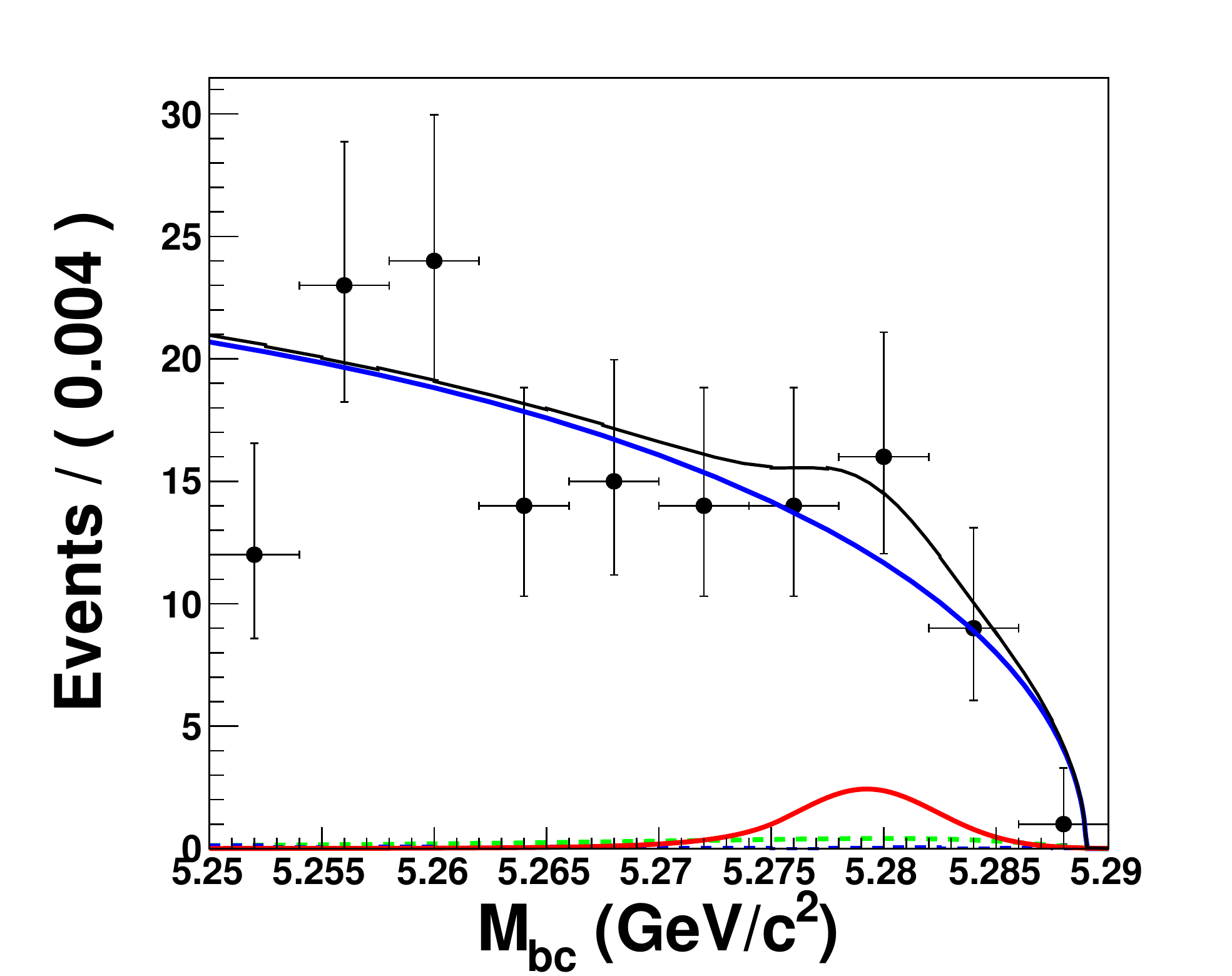}
 \includegraphics[width = 5cm]{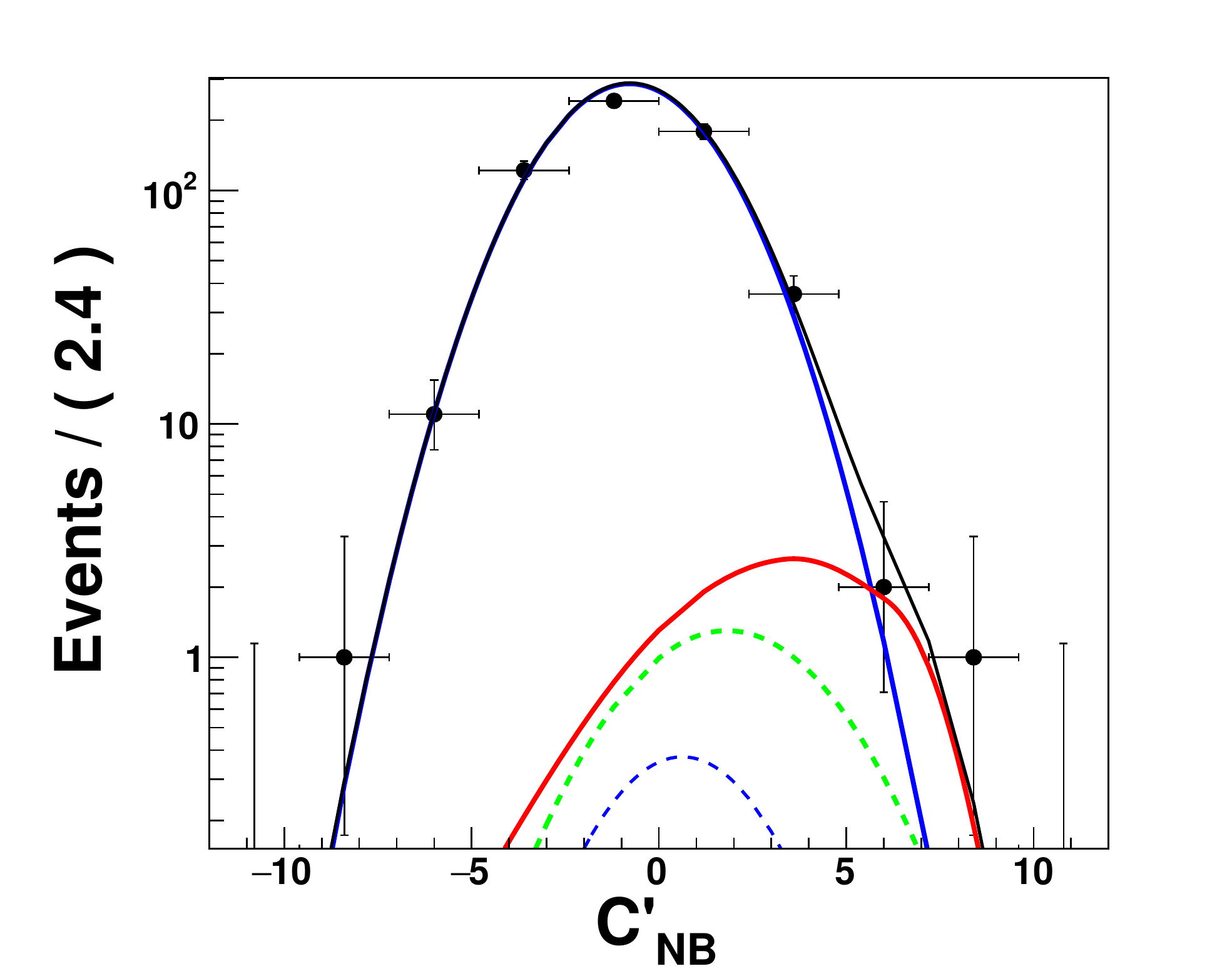}
}
 \subfigure[$B^{0}\to\eta_{3\pi}\eta_{3\pi}$]{  \includegraphics[width = 5cm]{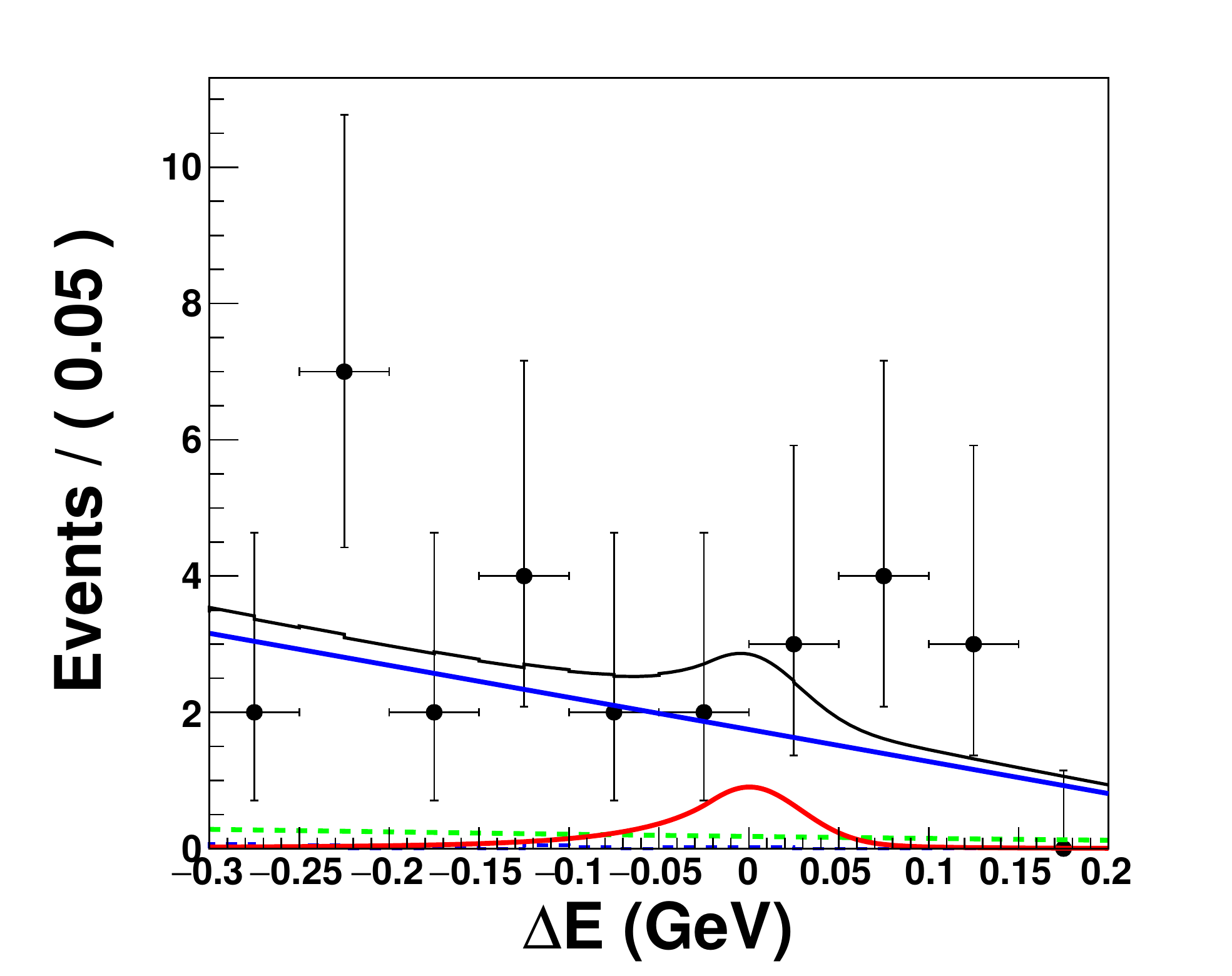} 
 \includegraphics[width = 5cm]{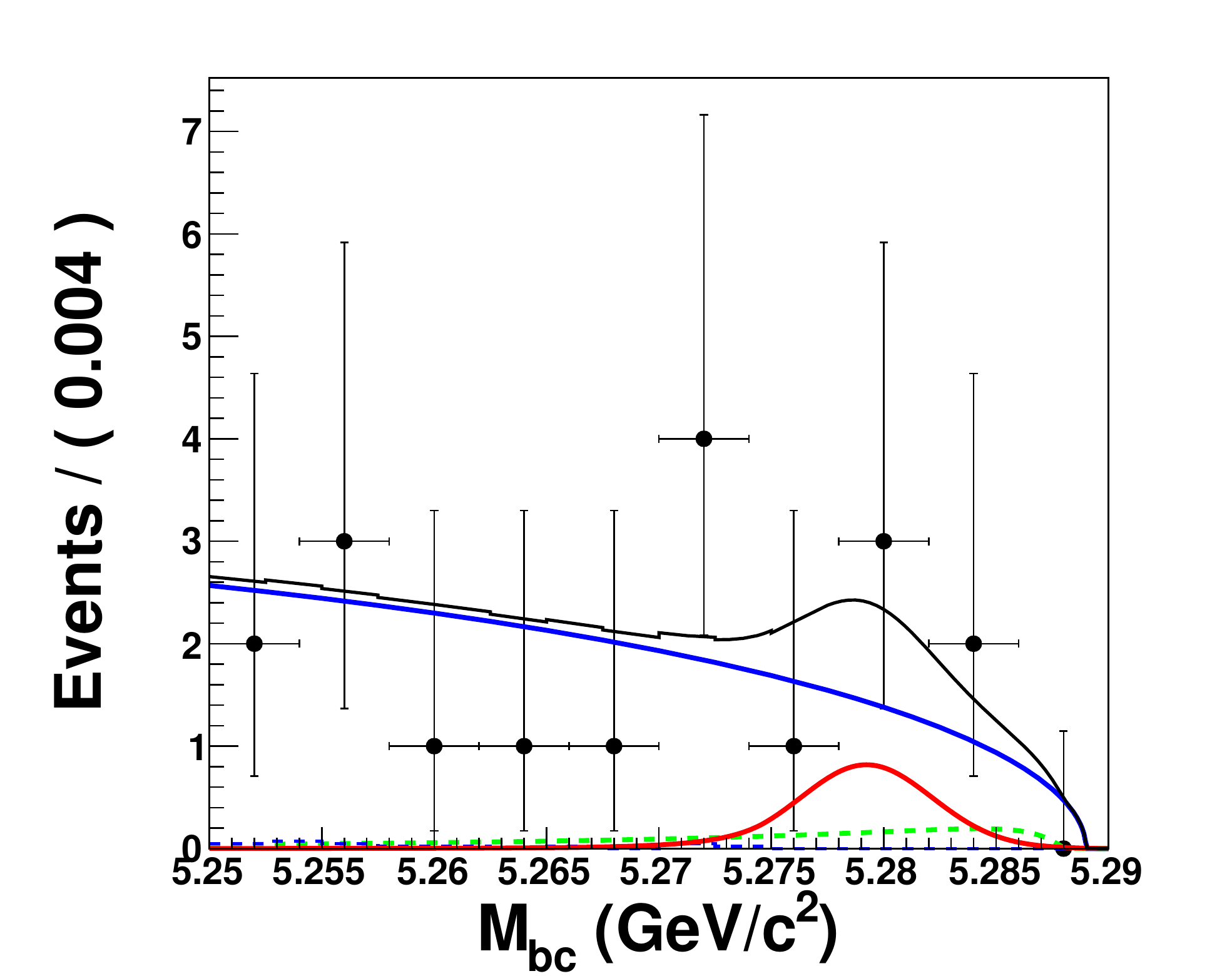}
 \includegraphics[width = 5cm]{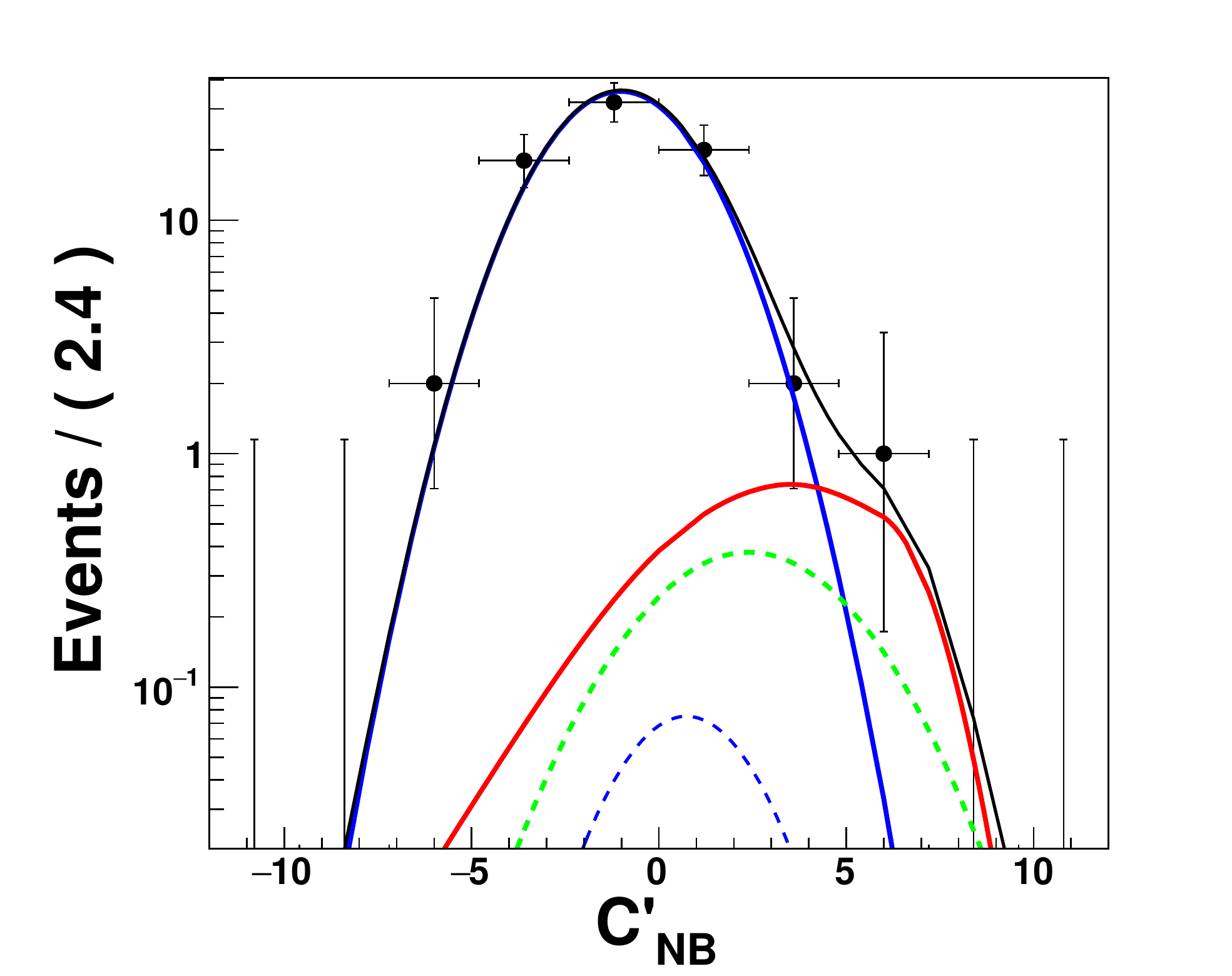}
}
 \caption{Signal-enhanced projections of the simultaneous fit: The points with the error bars are the real data, the black solid line is total PDF, the red solid line show the signal, the blue solid line represent the $q\bar{q}$, the green dashed line is $b \to u,d,s$, and the blue dashed line is $b\to c$ background.}
\label{fig:FitDataBox}
\end{figure*}

\section{FIT TO DATA}

We extract $23.6^{+8.1}_{-6.9}$, $9.2^{+3.2}_{-2.7}$ and $2.7^{+0.9}_{-0.8}$ signal events and  $3860.5^{+63.1}_{-62.4}$, $3779.7^{+62.0}_{-61.5}$ and $621.4^{+25.4}_{-24.8}$ continuum background events for $\eta_{\gamma\gamma}\eta_{\gamma\gamma}$ , $\eta_{\gamma\gamma}\eta_{3\pi}$ and $\eta_{3\pi}\eta_{3\pi}$, respectively.
The complete results of the ML fit are enumerated in Table \ref{table:conclusion}.
Figure \ref{fig:FitDataBox} shows the PDF and data distributions projected in the signal-enhanced region of $|\Delta E, M_{\rm bc}|<\pm 3\sigma$ and $C'_{NB} >$2.0, 2.0 and 1.5 for $\eta_{\gamma\gamma}\eta_{\gamma\gamma}$ , $\eta_{\gamma\gamma}\eta_{3\pi}$ and $\eta_{3\pi}\eta_{3\pi}$, respectively.
In the fit, the signal yield of each sub-decay mode (with index $k$) is written in terms of the common branching fraction as
\begin{eqnarray}
  n_{\mathrm{sig},k} = \mathcal{B}(B^{0}\to \eta \eta) \times N_{B\bar{B}} \times \epsilon_{\mathrm{rec},k} \times \prod \mathcal{B}_{\eta_{k}},
\end{eqnarray}
where $N_{B\bar{B}}$ is the number of $B\bar{B}$ pairs, $\epsilon_{\rm rec}$ is the signal efficiency obtained from MC simulation, $n_{\rm sig}$ is the number of signal events and $\prod \mathcal{B}_{\eta}$  is the product of the two $\eta$-decay branching fractions~\cite{PDG}.
The efficiency $\epsilon_{\rm rec}$ is corrected by the modest differences between data and MC in the particle identification efficiency as $\epsilon_{\rm PID}=0.966\pm0.008\,(0.965\pm0.008)$ per charged track in  $\eta_{\gamma\gamma}\,\eta_{3\pi}\,$$\eta_{3\pi}\eta_{3\pi}$) and in the retention rate of the $C_{\it NB}>-0.8$ cut as $\epsilon_{C_{\mathit{NB}}}=0.993^{+0.103}_{-0.097}$. 
The resulting branching fraction is $\mathcal{B}(B^{0} \to \eta \eta ) = (7.6^{+2.7}_{-2.3}) \times 10^{-7}$, where the error is statistical only. 

The significance $\mathcal{S}$ of the signal is defined as $\sqrt{-2\log{\mathcal{L}_{0}/\mathcal{L}_{\rm max}}}$, where $\mathcal{L}_{\rm max}\,$($\mathcal{L}_{0}$) is the likelihood value when the signal yield set to the measured signal yield$\,$(zero), corrected for the systematic errors by convolving the likelihood function with an asymmetric Gaussian distribution whose left and right variances equal the signal-yield systematic errors in Table \ref{table:uncertainty}.
The resulting significance of the branching fraction is 3.3 standard deviations above zero which provides the first evidence of this decay mode.

\begin{table}[t!]
\caption{\small Summary of results of the combined fit}
\begin{center}
\begin{tabular}{c c c  c}
\hline \hline
Sub-decay mode & $ \eta_{\gamma\gamma} \eta_{\gamma\gamma}$  &  $ \eta_{\gamma\gamma}  \eta_{3\pi}$ & $\eta_{3\pi} \eta_{3\pi}$   \\ \hline  
\multicolumn{4}{l}{Yields } \\
Signal & $23.6^{+8.1}_{-6.9}$ & $9.2^{+3.2}_{-2.7}$ & $2.7^{+0.9}_{-0.8}$\\
Continuum & $3860.5^{+63.1}_{-62.4}$ & $3779.7^{+62.0}_{-61.5}$ & $621.4^{+25.4}_{-24.8}$ \\
$b \to c$ (fixed)&5.9 & 5.9 & 2.2 \\
$b \to u, d, s $ (fixed) & 27.4 &17.8 & 4.5 \\
$B^{0}\to \eta \pi^{0}$ (fixed)& 1.4 & 0.1 & - \\
\multicolumn{4}{l}{Efficiency} \\ 
$\epsilon_{\rm rec}(\%)$ & 26.3 & 17.8 & 8.9 \\
$\prod \mathcal{B}_{\eta}(\%)$   & 15.5 & 8.9 & 5.1 \\\hline
$\mathcal{B}(\times 10^{-7})$ & \multicolumn{3}{c}{$7.6^{+2.7 +1.4}_{-2.3 -1.6}$}\\
$\mathcal{B}$ significance $\mathcal{S}(\sigma)$ & \multicolumn{3}{c}{3.3}\\
\hline\hline
\end{tabular}
\end{center}

\label{table:conclusion}
\end{table}

\section{SYSTEMATIC UNCERTAINTY ESTIMATION}
The systematic uncertainties in the branching fraction are listed in Table \ref{table:uncertainty}.
The uncertainty due to the fixed parameters in the PDF is estimated by varying each, one by one, according to its statistical uncertainty. Deviations from the original fit are added in quadrature. 
We vary the bin height for all histogram PDFs by the bin's statistical error and repeat the fit. The resulting changes are added in quadrature and the result is taken as the systematic uncertainty. 
The uncertainty due to calibration factors are evaluated in a similar manner.
The uncertainty due to the fixed fractions of misreconstructed events and fixed yields are calculated by varying them by $\pm50\%$.

We determine the uncertainty due to the slightly different continuum suppression efficiencies for $C_{\mathit{NB}} = -0.8$ in data and MC by using the $B^{0}\to\overline{D}{}^{0}\eta$ control sample. 
The systematic uncertainty due to the charged-track reconstruction efficiency is estimated to be $0.35\%$ per track by using a partially reconstructed $D^{*+}\to D^{0}(K^{0}_{S}\pi^{+}\pi^{-})\pi^{+}$ events. 
An uncertainty of $0.8\%$ per track is assigned due to PID criteria. 
The uncertainty in the reconstruction efficiency of each $\eta\to\gamma\gamma$ or $\pi^{0}\to \gamma \gamma$ decay is $3\%$~\cite{int28}. 

We assign systematic uncertainties of $0.5\%$ and $1.2\%$, respectively, for the branching fractions of $\eta \to \gamma \gamma$ and $\eta \to \pi^{+}\pi^{-}\pi^{0}$~\cite{PDG}.
The uncertainty in the efficiency $\epsilon$ due to the limited signal MC statistics is $0.3\%$ and the uncertainty due to the number of $B\bar{B}$ pairs is $1.3\%$.

\begin{figure}[h!]
 \centering
\subfigure[ ]{ \includegraphics[width = 4cm]{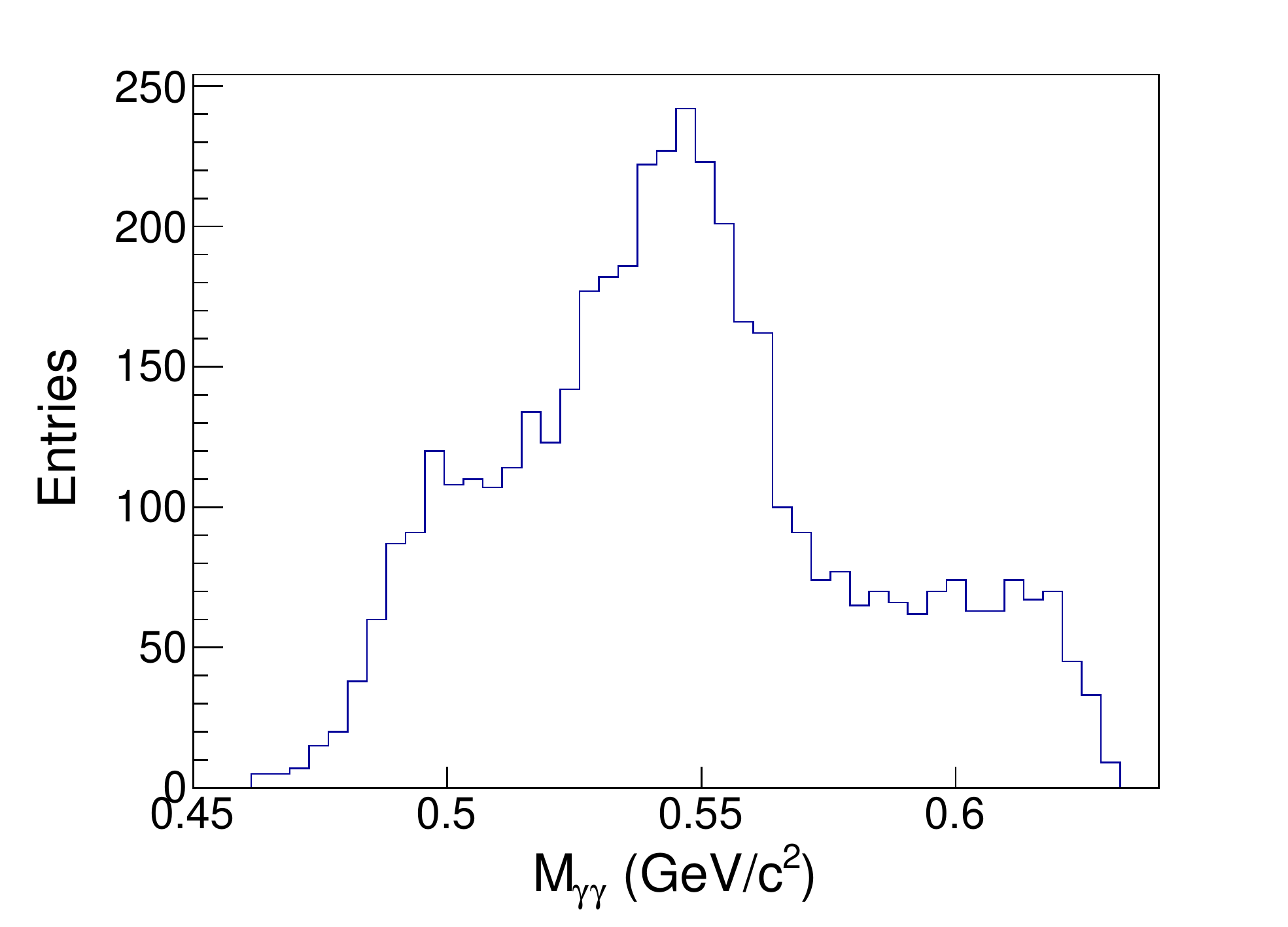}}
 \subfigure[ ]{ \includegraphics[width = 4cm]{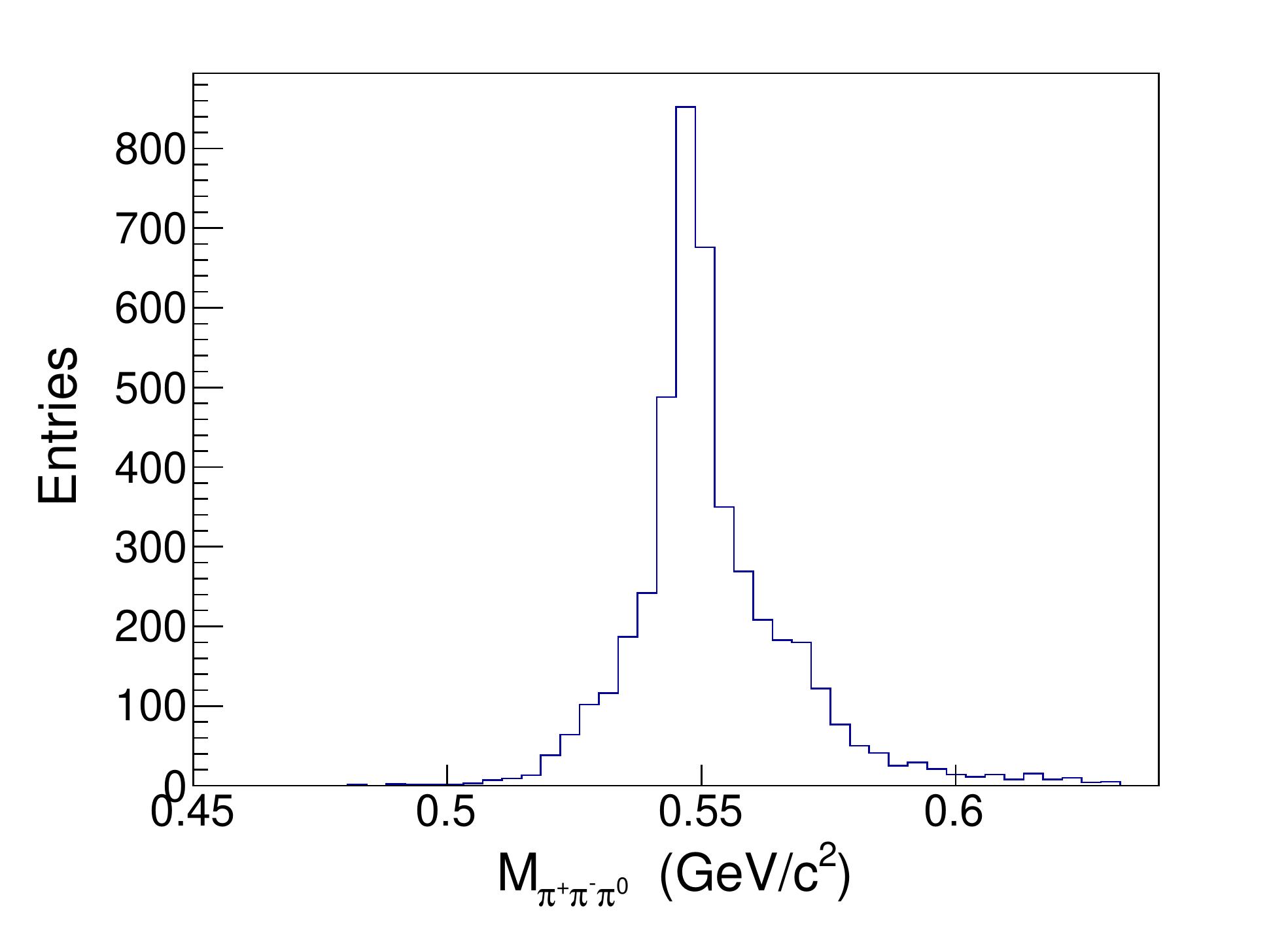}}
 \caption{Distribution of $M_{\gamma\gamma}$(a) and $M_{\pi^{+}\pi^{-}\pi^{0}} $(b) invariant masses for events passing all selection requirements, except for  $M_{\gamma\gamma}$ or  $M_{\pi^{+}\pi^{-}\pi^{0}} $}
\label{fig:etaSB}
\end{figure}

In order to check for potential non-resonant $B^{0} \to \eta \gamma \gamma$, $B^{0} \to \gamma \gamma \gamma \gamma $, $B^{0} \to \gamma \gamma \eta_{\pi^{+}\pi^{-}\pi^{0}}$, $B^{0} \to \eta_{\gamma \gamma} \pi^{+}\pi^{-}\pi^{0}$, $B^{0} \to \gamma \gamma \pi^{+}\pi^{-}\pi^{0}$, $B^{0}\to\eta_{\pi^{+}\pi^{-}\pi^{0}} \pi^{+}\pi^{-}\pi^{0}$ and $B^{0} \to \pi^{+}\pi^{-}\pi^{0} \pi^{+}\pi^{-}\pi^{0}$ contamination, we relax the $\eta$ mass requirement; the invariant mass distributions are shown in Fig. \ref{fig:etaSB}.
We choose the $\eta_{\gamma\gamma}\,$($\eta_{3\pi}$) mass-sideband region as $0.45-0.48\,\textrm{GeV}/c^{2}$ or $0.58-0.63\,\textrm{GeV}/c^{2}$ ($0.45-0.52\, \textrm{GeV}/c^{2}$ or $0.57-0.63\,\textrm{GeV}/c^{2}$) and repeat the fitting procedure.
We measure the branching fraction of the non-resonant contribution to be $(0.02^{+0.09}_{-0.05}) \times 10^{-6}$, consistent with zero.
 The positive uncertainty is assigned as the negative systematic uncertainty. 

\begin{table}[t!]
\caption{Summary of systematic uncertainties$\,$($\%$) on branching fraction. The list in upper section is associated with fitting for the signal yields and are included in the signal significance.}
\begin{center}
\begin{tabular}{c c}
\hline 
Source & Uncertainty ($\%$) \\ \hline \hline 
Fit bias  &$_{-1.7}^{+0.0}$\\
PDF Modeling  &4.8 \\
Histogram PDF  &1.2\\
Fraction of SCF  &2.3\\
Fixed yields  &6.4\\
Calibration factors  &10.2 \\ \hline
$C_{\mathit{NB}}$ efficiency  &$_{-9.7}^{+10.3}$\\
Non-resonance contribution  &$_{-11.8}^{+0.0}$ \\
MC statistics & 0.3\\
PID efficiencies & 2.1\\
Tracking efficiency & 1.0 \\
$\pi^{0}, \eta \to \gamma \gamma$  & 6.0\\
$\mathcal{B}(\eta \to \gamma \gamma)$ & 0.8 \\
$\mathcal{B}(\eta \to \pi^{+} \pi^{-} \pi^{0})$  & 1.8 \\
The number of $B\bar{B}$ events. & 1.3\\	
\hline 
Total  &$_{-21.4}^{+18.2}$ \\
\hline\hline
\end{tabular}
\end{center}
\label{table:uncertainty}
\end{table}

\section{CONCLUSION}

In summary, we have conducted a measurement of the branching fraction of the decay $B^{0}\to \eta \eta$. 
We obtain 
\begin{eqnarray}
\mathcal{B}(B^{0} \to \eta \eta ) = (7.6^{+2.7 +1.4}_{-2.3 -1.6}) \times 10^{-7} \nonumber ,
\end{eqnarray}
where the first uncertainty is statistical and the second is systematic. 
The significance of this result is 3.3 standard deviations above zero, which provides the first evidence for this decay. 
The measured branching fraction is in good agreement with the theoretical expectations~\cite{int12, int13, int14}.

\section{ACKNOWLEDGMENTS}

We thank the KEKB group for the excellent operation of the
accelerator; the KEK cryogenics group for the efficient
operation of the solenoid; and the KEK computer group,
the National Institute of Informatics, and the 
PNNL/EMSL computing group for valuable computing
and SINET4 network support.  We acknowledge support from
the Ministry of Education, Culture, Sports, Science, and
Technology (MEXT) of Japan, the Japan Society for the 
Promotion of Science (JSPS), and the Tau-Lepton Physics 
Research Center of Nagoya University; 
the Australian Research Council;
Austrian Science Fund under Grant No.~P 22742-N16 and P 26794-N20;
the National Natural Science Foundation of China under Contracts 
No.~10575109, No.~10775142, No.~10875115, No.~11175187, No.~11475187
and No.~11575017;
the Chinese Academy of Science Center for Excellence in Particle Physics; 
the Ministry of Education, Youth and Sports of the Czech
Republic under Contract No.~LG14034;
the Carl Zeiss Foundation, the Deutsche Forschungsgemeinschaft, the
Excellence Cluster Universe, and the VolkswagenStiftung;
the Department of Science and Technology of India; 
the Istituto Nazionale di Fisica Nucleare of Italy; 
the WCU program of the Ministry of Education, National Research Foundation (NRF) 
of Korea Grants No.~2011-0029457,  No.~2012-0008143,  
No.~2012R1A1A2008330, No.~2013R1A1A3007772, No.~2014R1A2A2A01005286,No.~2014R1A2A2A01002734, No.~2015R1A2A2A01003280 , No. 2015H1A2A1033649;
the Basic Research Lab program under NRF Grant No.~KRF-2011-0020333,
Center for Korean J-PARC Users, No.~NRF-2013K1A3A7A06056592; 
the Brain Korea 21-Plus program and Radiation Science Research Institute;
the Polish Ministry of Science and Higher Education and 
the National Science Center;
the Ministry of Education and Science of the Russian Federation and
the Russian Foundation for Basic Research;
the Slovenian Research Agency;
Ikerbasque, Basque Foundation for Science and
the Euskal Herriko Unibertsitatea (UPV/EHU) under program UFI 11/55 (Spain);
the Swiss National Science Foundation; 
the Ministry of Education and the Ministry of Science and Technology of Taiwan;
and the U.S.\ Department of Energy and the National Science Foundation.
This work is supported by a Grant-in-Aid from MEXT for 
Science Research in a Priority Area (``New Development of 
Flavor Physics'') and from JSPS for Creative Scientific 
Research (``Evolution of Tau-lepton Physics'').

%

\end{document}